\documentclass[reprint,
 amsmath,amssymb,
 aps]{revtex4-1}

 \usepackage{hyperref}
\hypersetup{colorlinks=true, citecolor=blue, urlcolor=blue, linkcolor=blue}

\usepackage{color}
\usepackage{graphicx}

\usepackage{cancel}
\usepackage[normalem]{ulem}

\usepackage{amsmath}

\usepackage[utf8x]{inputenc}

\begin{document}

\title[{Relevance of microscopic activation rules}]{Apparently similar neuronal dynamics may lead to different collective repertoire}

\author{Margarita M. Sánchez Díaz$^{1,3}$, Eyisto J. Aguilar Trejo$^{1,2,3}$, Daniel A. Martin$^{1,2,3}$,
Sergio A. Cannas$^{2,4}$,
Tomás S. Grigera$^{2,5,6}$,
Dante R. Chialvo$^{1,2,3}$
}

\affiliation{$^1$ Center for Complex Systems and Brain Sciences (CEMSC$^3$), Universidad Nacional de San Mart\'in, Campus Miguelete,
25 de Mayo y Francia, (1650), San Mart\'in, Buenos Aires, Argentina.}

\affiliation{$^2$ Consejo Nacional de Investigaciones Científicas y T\'ecnicas (CONICET),
Godoy Cruz 2290, (1425), Buenos Aires, Argentina.}

\affiliation{$^3$ Instituto de Ciencias F\'isicas (ICIFI), CONICET and Universidad Nacional de San Mart\'in
25 de Mayo y Francia, (1650), San Mart\'in, Buenos Aires, Argentina.}

\affiliation{$^4$ Instituto de Física Enrique Gaviola (IFEG-CONICET), Facultad de Matemática Astronomía Física y Computación,
Universidad Nacional de Córdoba, 5000 Córdoba, Argentina.}

\affiliation{$^5$ Instituto de F\'isica de Líquidos y Sistemas Biológicos (IFLySiB), CONICET and Universidad Nacional de La Plata, Calle 59 no.\ 789, B1900BTE,
La Plata, Buenos Aires, Argentina.}

\affiliation{$^6$ Departamento de F\'isica, Facultad de Ciencias Exactas, Universidad Nacional de La Plata, 1900, La Plata, Buenos
Aires, Argentina.}

\begin{abstract}
This report is concerned with the relevance of the microscopic rules, that implement individual neuronal activation, in determining the collective dynamics, under variations of the network topology. To fix ideas we study the dynamics of two cellular automaton models, commonly used, rather in-distinctively, as the building blocks of large scale neuronal networks. One model, due to Greenberg \& Hastings, (GH) can be described by evolution equations  mimicking  an integrate-and-fire process, while the other model, due to Kinouchi \& Copelli, (KC) represents an abstract branching process, where a single active neuron activates a given number of postsynaptic neurons according to a prescribed ``activity'' branching ratio.  Despite the apparent similarity between the local neuronal dynamics of the two models, it is shown that they exhibit very different collective dynamics as a function of the network topology. The GH model shows qualitatively different dynamical regimes as the network topology is varied, including transients to a ground (inactive) state, continuous and discontinuous dynamical phase transitions.  In contrast, the KC  model only exhibits a continuous phase transition, independently of the network topology. These results highlight the importance of paying attention to the microscopic rules chosen to model the inter-neuronal interactions in large scale numerical simulations, in particular when the network topology is far from a mean field description. One such case is the extensive work being done in the context of  the Human Connectome, where a wide variety of types of models are being used to understand the brain collective dynamics.
\end{abstract}


\maketitle

\section{Introduction}

The animal brain is composed by billions of neurons, which interact with each other through thousands of synapses per neuron. The results of such interaction is the emergence of complex spatio-temporal patterns of neuronal activity supporting perception, action and behavior. A recent proposal considers the brain as a network of neurons  poised near a dynamical transition \cite{bak, chialvo2004critical, chialvo2010emergent, mora2011biological}, a view which is supported by experimental results gathered from animals both \emph {in vitro}\cite{BeggsYPlenz} and \emph {in vivo} \cite{Tiagotesis} as well as from whole brain neuroimaging human experiments \cite{expert,FraimanChialvo2012,tagliazucchi2012}.

The  potential existence of critical phenomena in the brain motivated during the last decade the study of mathematical models to better explore the large-scale brain dynamics. A distinctive difference between the diversity of models is at the microscopic level. Some models consist of networks of simplified neurons, in which neurons themselves are represented by a wide variety of approaches, ranging from 2-state particles \cite{odor2016critical} through  discrete cellular automatons \cite{Haimovici2013,Zarepour,rocha1,Moosavi}, branching processes \cite{Kinouchi2006,Viola2014,Tiagotesis,shew2009neuronal,HaldemanBeggs}, neural masses \cite{neuralmass},  coupled-maps \cite{chialvo1995,rulkov,Girardi-Schappo,cmlreview}, coupled Kuramoto oscillators \cite{ kuramoto_refs} up to detailed equations describing the evolution and spiking of the membrane potential \cite{levina,izhikevich2003simple,Poil}. Thus, a natural question arises on how relevant may the microscopic process used to represent the individual neuronal dynamics be, and how they affect the dynamical collective repertoire exhibited by the network.

When focusing on collective properties, it is of course reasonable to seek minimal models which, even orphan of realistic microscopic rules, may reproduce relevant macroscopic behavior.  However, it is not straightforward to determine in principle how general this assumption can be in the case of neuronal networks. Our point is that, even though the use of realistic microscopic dynamics is not necessarily a prerequisite to correctly describe universal macroscopic properties, \emph{microscopic rules do matter} and eventually can lead to different universal behavior. As a clarifying metaphor consider the Ising model.  It is well known that algorithms with unrealistic non-local moves (so-called cluster algorithms \cite{clusteralgorithms}) can correctly describe the static critical behavior.  However, if the dynamic rule did not follow detailed balance, then the modified dynamics would fail to reproduce equilibrium behavior, even if it could reproduce some sort of critical dynamics.  And of course, even with detailed balance, the dynamical universality is altered by the non-local rule. A similar correspondence among microscopic rules and system's dynamics appears when modeling brain dynamics, which is rarely considered, thus some extrapolations to real brain dynamics taken from numerical simulations in the current literature, may be hampered by the limitations of the microscopic details of the neuronal models employed. We are purposely not considering here a large chapter of models that include synaptic plasticity. 

In this article, we illustrate the problem by studying the dynamics of two apparently similar neural network models: that of Greenberg \& Hastings \cite{Greenberg} as described in \cite{Haimovici2013,Zarepour} and that of Kinouchi \& Copelli \cite{Kinouchi2006}.  The main difference between these two models is related to the microscopic rule that propagates the activity: the first proposes (as many others of the same kind) a neuronal interaction rule that depends on the state of its presynaptic neighbors, while the second introduces a rule that, regardless of the number of connections, maintains a prescribed branching of activity on the target neurons. At first sight the differences seem innocent-looking, but as it will be shown, they lead to completely different  behavior of the network: the first model exhibits continuous or discontinuous phase transitions depending on the network topology, while the second is completely insensitive to it. 

We remark from the outset that the article' aim is not to criticize any given model in particular, but to call the attention on the consequences of using them ignoring the limitations of the model's original formulation, including possible misinterpretations. The article is organized as follows: in Sec.~\ref{Model} we describe for both models the observables that will be used to characterize the dynamical regimes, in Sec.~\ref{Results} we show the results of the numerical simulations, in Sec.~\ref{Discussion} we discuss present results in the context of recent research and we summarize the conclusions.

\section{Network, models and observables}
\label{Model}

\subsection{The interaction network}

Both neuronal models are studied on an undirected Watts-Strogatz small-world network \cite{watts1998collective} with average connectivity $\langle k \rangle$ and rewiring probability $\pi$.  The network is constructed as usual \cite{watts1998collective} by starting from a ring of $N$ nodes (always $N=20000$ in this report), each connected symmetrically to its $\langle k\rangle /2$ nearest neighbors; then each link connecting a node to a clockwise neighbor is rewired to a random node with probability $\pi$, so that average connectivity is preserved.  The rewiring probability is a measure of the disorder in the network: for $\pi=0$ the network is circular and perfectly ordered, while for $\pi=1$ it becomes completely random.
\begin{figure}[h]
\centering
\includegraphics[width=.7\linewidth]{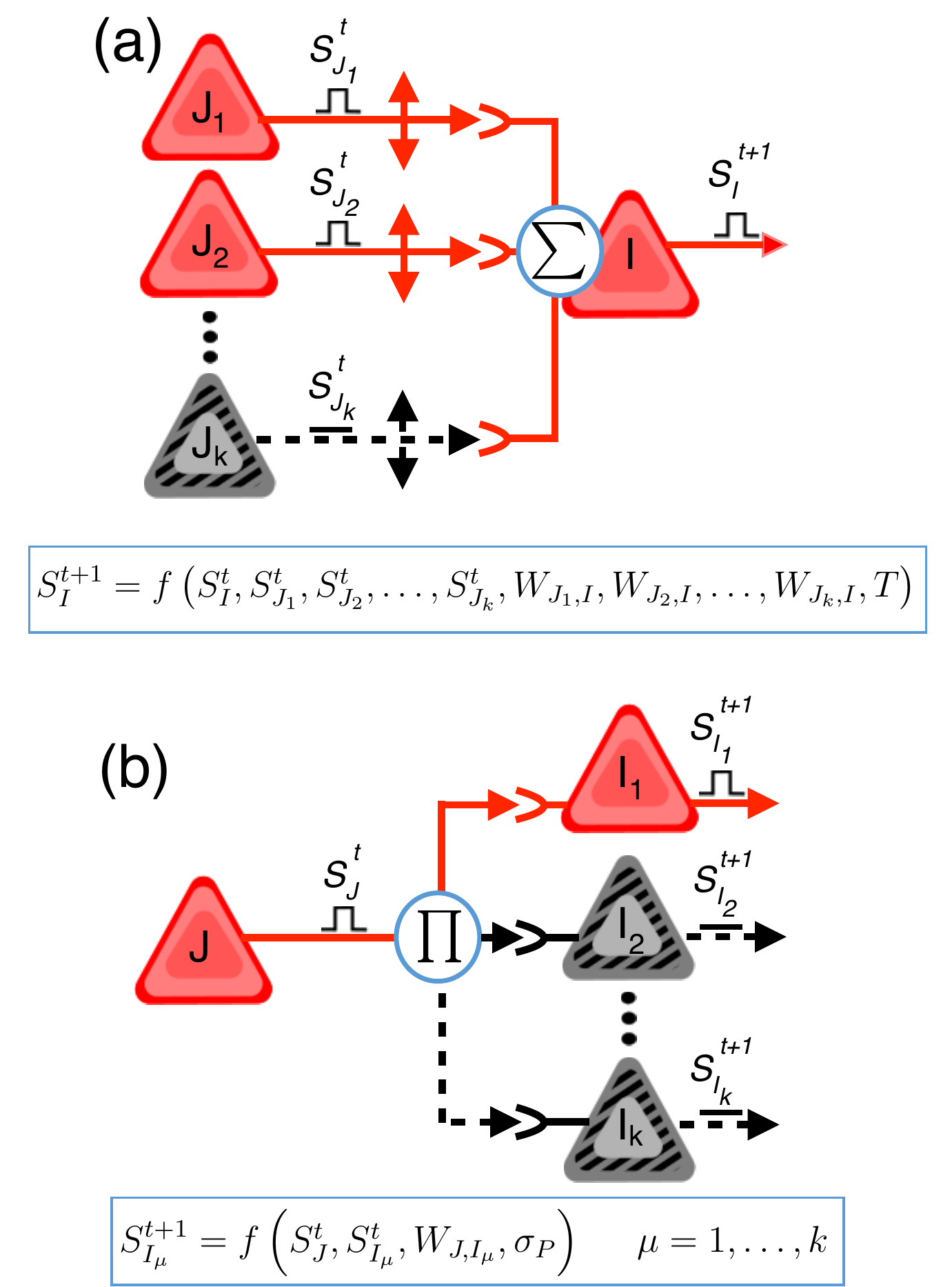}
\caption{{Rule for the propagation of activity in both models.}  Panel (a): In the GH model,  a \emph{given neuron $I$} will become active at time $t+1$, if the contribution of all active presynaptic neurons (here $J_1$ and $J_2$), weighted by the interaction $W_{I,J}$ surpasses the threshold $T$.  At each time step, this update is repeated for \emph{all quiescent neurons $I$}. Panel (b): In the KC model \emph{ a given active neuron $J$} will activate at time step $t+1$ a given number of $I$ neurons depending on the $W_{I,J}$ and $\sigma_p$ value. Note that $\sigma_p$ normalizes such probability by  the number of interactions $\langle k \rangle$. At each time step, this update is repeated for \emph{all active neurons $J$}. In both panels, triangles represent neurons and lines synaptic interactions. Red filled triangles denote\sout{s} active neurons and gray dashed ones inactive neurons, at times $t$ and $t+1$.
}\label{Fig1}
\end{figure}
In both models neurons are represented as nodes on a weighted undirected random graph with an associated discrete state variable, $S_i=0,\ldots,n$, where $i=1,\ldots,N$ identifies the node and $S_i=\{0,\ldots,n\}$.  State 0 represents a quiescent (but excitable) neuron, 1 is the active state, and $2\ldots n$ are refractory states.  The links of the graphs are represented by the $N\times N$ connectivity matrix $W$.  Nonzero matrix elements indicate the presence of a link with a given weight.  Weights are positive reals, so $W_{ij}\ge0$, and the connectivity matrix is symmetric, $W_{ij}=W_{ji}$, since the graph is undirected. In this context symmetric connections need to be interpreted as two connections between any pair of nodes. Neither the connectivity nor the weights depend on time (i.e., we consider quenched disorder). The dynamical evolution is given by a discrete-time Markov process in which all sites are simultaneously updated, and with transition probabilities for each site given by the expressions below for each model.

\subsubsection{GH model:}
This model was introduced by Greenberg \& Hastings \cite{Greenberg} to mimic the excitable dynamics generically observed in neurons, forest fires, cardiac cells, chemical reactions and epidemic propagation. In the context of brain dynamics it was used recently  by Haimovici \emph{et al.} \cite{Haimovici2013}. Here we follow closely the implementation of Zarepour \emph{et al.} \cite{Zarepour}.   It is a cellular automaton endowed of the three states common to excitable dynamics: quiescent, active and refractory state, and the dynamics of site $i$ is updated by
\begin{subequations}
\begin{align}
  P_{i,0\to1} &= 1- [1-r_1] \left[ 1- \Theta \left(\sum_{j=1}^{k_{in,i}} W_{j,i} \delta_{S_j,1}-T \right)\right], \label{eq:act-modA} \\
  P_{i,1\to2} & =1, \\
  P_{i,2\to0} & = r_2,
\end{align}
\end{subequations}
where $P_{i,a\to b}$ is the probability that site $i$ will transition from state $a$ to state $b$, at time $t+1$, $S_i$ is computed at time $t$, and the sum is performed over all $j$ targeting  $i$ {and $k_{in,i}$ is the in-degree of node $i$}. $\Theta(x)$ is Heaviside's step function [$\Theta(x)=1$ for $x\ge0$ or 0 otherwise], $\delta_{i,j}$ is Kronecker's delta, and $r_1$, $r_2$ and $T$ are control parameters which are set equal to all sites in the present work.  Thus an active site always turns refractory in the next time step, and a refractory site becomes quiescent with probability $r_2$.  The probability for a quiescent site to become active is written as 1 minus the product of the probabilities of \emph{not} becoming active through the different mechanisms at work.  In this model there are only two activation mechanisms: \emph{spontaneous activation}, which occurs with a small probability $r_1$, or \emph{transmitted activation,} which occurs deterministically whenever the sum of the weights of the links connecting $i$ to its active neighbors exceeds a threshold $T$ (see Panel (a) of Fig. \ref{Fig1}). The non-null weights are drawn from an exponential distribution, $p(W_{ij}=w)=\lambda e^{-\lambda w}$, with $\lambda=12.5$ chosen to mimic the weight distribution of the human connectome \cite{Zarepour}.  For the simulations described here, we use $r_1=0.001$, $r_2=0.3$ as in previous work \cite{Haimovici2013,Zarepour} which remain fixed in all simulations, while $T$ is used as control parameter.

\subsubsection{KC model:}

This model was introduced by Kinouchi \& Copelli \cite{Kinouchi2006} to  show that a (Erd\"os–Renyi undirected) network of excitable elements has its sensitivity and dynamic range maximized at the critical point of a non-equilibrium phase transition. The model resembles a branching process \cite{Branching,BranchingProcessExponents} in which the transition probabilities for neuron $i$ at time $t+1$ are:

\begin{subequations}
\begin{align}
P_{i,0\to1} & = 1- [1-r_1] \prod_{j=1}^{k_{out,i}} [1-p W_{ji} \delta_{S_j,1}],\label{eq:act-modB} \\
 P_{i,1\to2} & =1, \\
 P_{i,2\to3} & =1, \\
  \vdots \notag \\
  P_{i,n\to0} & = 1.
\end{align}
\end{subequations}

$S_j$ is evaluated at $t$, and the product is taken over all neurons $j$ pointing to $i$.  The interaction rule in Eq. \ref{eq:act-modB} contains two parameters: $r_1$ which (as in the GH model) determines the spontaneous activity of any inactive neuron and $\sigma_p$ which acts as a control parameter. (see Panel (b) of Fig. \ref{Fig1}). This rule makes the main difference with the GH model:  here an active site $j$ activates {\em a given number of neighbors $i$}  with probability $\sigma_p W_{ji}$. If the variance of the chosen values for $k_{out}$ and $W$ are relatively small, (as in ref. \cite{Kinouchi2006}) and $\langle k \rangle$ is relatively large, each active neuron will excite, on average $\sigma_p \doteq (\langle k \rangle -1) p/2$ neurons. Thus $\sigma_p$ represents the desired branching ratio. It is known that, for a wide variety of conditions,  critical dynamics is expected  for $\sigma_p\simeq 1$ \cite{Branching}.
The interaction matrix is symmetric, $W_{ij}=W_{ji}$ following \cite{Kinouchi2006}, and non-null elements are taken uniformly from $[0,1]$. Similar to the GH model, an active site always becomes refractory, but instead of recovering randomly, here it becomes quiescent deterministically after $n-1$ time steps. We note that this difference has no relevance for the present analysis.

For the simulations, values of $r_1=0.001$, and $n=4$ (i.e.,\ a fixed refractory period of 3 steps) are chosen, which remain fixed in all simulations.  In passing, please notice that the interaction rule in the KC model is entirely stochastic and that neurons behave independently (as long as spontaneous activity is relatively low as dictated by the value of $r_1$ used here). Additional details can be found in Costa et al. \cite{Costa} and Campos et al. \cite{Campos}. The numerical implementation of the KC model admits a few variations  which, nonetheless, does not change the present results (see Supplemental Material \cite{repository}).

\subsection{Observables}

To describe the state of the network, for both models, we define an order parameter $f_S(t)$ which corresponds to the fraction of active neurons at time $t$,
\begin{equation}
f_S(t)= {1 \over N} \sum_i \delta_{S_i(t),1}.
\end{equation}
After any transient dies out, we also compute its variance, $\sigma_{f_S}^2=\langle f_S^2\rangle - \langle f_S\rangle ^2$, where  $\langle ...\rangle$ is a time average.

For the purposes of the present work, it is of particular interest the behavior of the (normalized) connected autocorrelation of the order parameter $f_S$,
\begin{equation}
AC(\Delta t) = {1\over \sigma_{f_S}^2} \Bigl\langle \bigl(f_S(t) - \langle f_S\rangle \bigr)  \times   \bigl( f_S(t+ \Delta t) - \langle f_S\rangle \bigr) \Bigr\rangle.
\end{equation}
which estimates the linear correlation between the network state at times $t$ and $t+\Delta t$,  with $AC(\Delta t) \simeq 1$ for highly correlated consecutive configurations and $AC(\Delta t) \simeq 0$ when the configurations quickly decorrelate.  It is known that the autocorrelation function is sensitive to the different dynamical regimes: close to a continuous phase transition, the dynamics undergoes critical slowing down, which implies that the autocorrelation function decays slower than in the supercritical or subcritical state \cite{Chialvo2020Control}.  For discontinuous phase transitions, a similar effect takes place at the spinodal points \cite{loscar_nonequilibrium_2009}.  Here we focus on the autocorrelation at $\Delta t=1$, $AC(1)$, also called first correlation coefficient, which has been shown to have a maximum at the transition point \cite{Chialvo2020Control}.

\subsection{Parametric exploration}
In this work we are interested in exploring the extent of the dynamical repertoire that each model is able to exhibit under a very wide range of: 1) neuronal dynamics and 2) topology of the underlying network. Thus, we proceed to scan the control parameter of the given neuron model for different  network topologies (by varying $\langle k \rangle$ and $\pi$). This implies to explore three parameters while classifying the dynamical regimes observed.

To identify and classify the dynamical regimes we track the behavior of $AC(1)$ as the  control parameter ($T$ or $\sigma_p$) is increased and decreased. This is repeated for each combination of  network parameters $\langle k \rangle$ and $\pi$. The simulations start at $T_0$ (or $\sigma_0$) (using a random initial condition for each neuron) and then it is increased  by $\Delta T$ (or $\Delta \sigma$)  after a given number of steps, up to a final value $T_F$  (or $\sigma_F$), without resetting the neuron states when changing the value of the control parameter.
This parametric exploration allows us to determine the full repertoire of dynamical regimes which can emerge from the microscopic activity propagation rules acting on a given network topology.
\begin{figure}[h!]
\centering
\includegraphics[width=.7\linewidth]{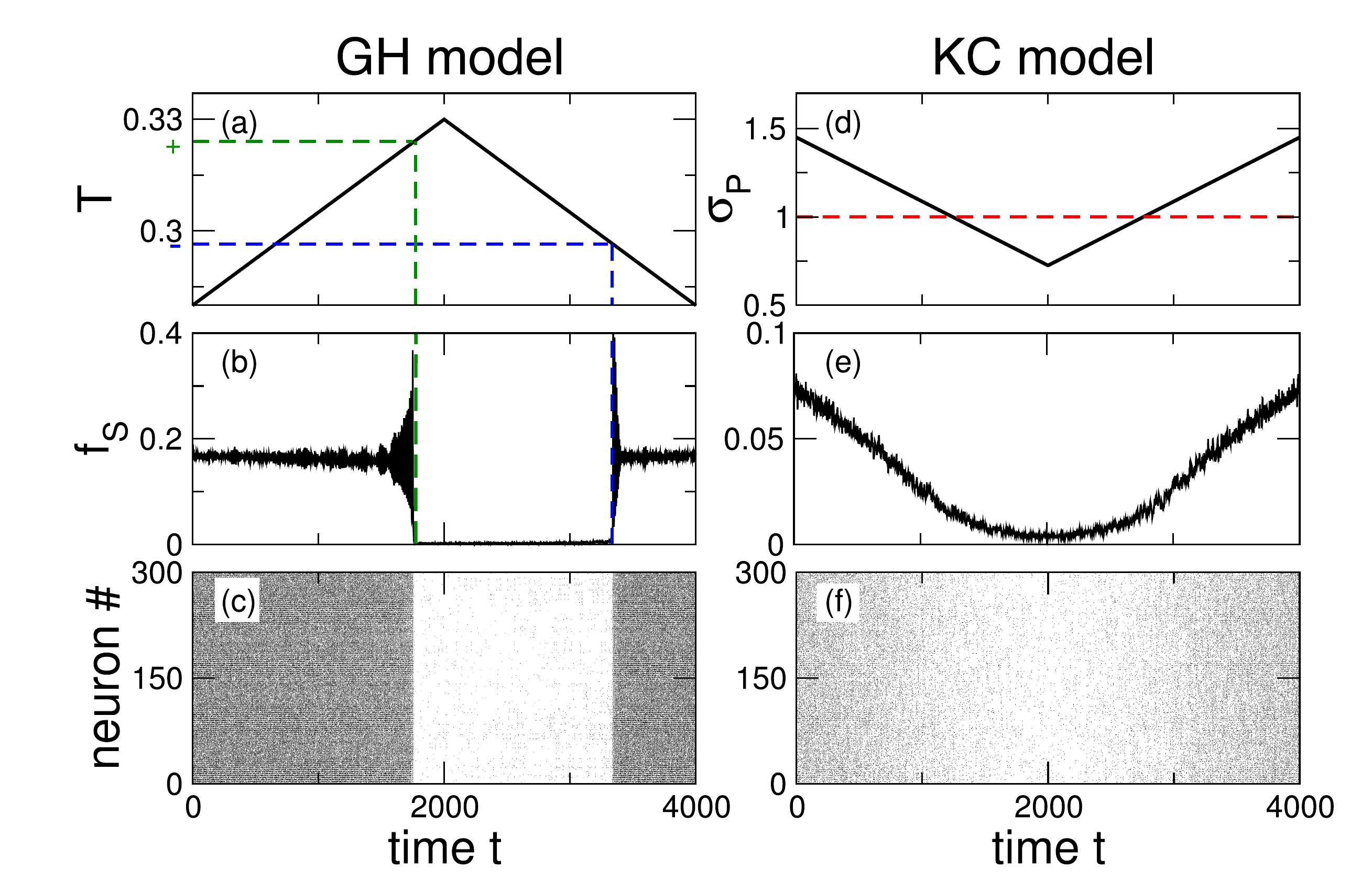}
\caption{Examples of the evolution of the activity as the respective control parameters ($T$ or $\sigma_p$) is slowly varied. Left panels (a-c) correspond to the GH model and right panels (d-f) to  the KC model. Top panels: value of the control parameters ($T$ in (a) and $\sigma_p$ in (d)) as a function of time $t$.  Center panels: order parameter  $f_S(t)$ for both models as a function of time.  Bottom panels: raster plots of 300 selected neurons.  Dashed lines in panels (a) and (b) mark the approximate values of $T_{\pm}$, and their respective times.  In both cases $\langle k \rangle=30$, $\pi=0.6$. For the GH model, $\Delta T=\pm0.000025$, and for the KC model $\Delta \sigma=\pm0.00036$. } \label{Fig2}
\end{figure}

\section{Results}
\label{Results}
\subsection{Characterizing the transitions}
Now we  proceed to describe how  the dynamical repertoire of each model is determined from parametric exploration. An example is presented in Fig.~\ref{Fig2} where panels on the left correspond to results obtained from the GH model and those on the right from the KC model. The figure shows that, as expected, the rate of activity changes as a function of its control parameter, but already demonstrating an important difference between the dynamical regimes exhibited by the two models.  For this particular choice of topology, $\langle k \rangle=30$ and $\pi=0.6$, the GH model undergoes a discontinuous transition demonstrated by the abrupt change in $f_s$ (also noted in the appearance of the raster plot) and the presence of hysteresis. In contrast, in response to similar parametric scan, the KC model exhibits a continuous transition and does not show hysteresis.  In addition, it is important to note that the GH model shows a large increase in the variability of the order parameter $f_s$ near the transition (see panel b), meanwhile the variance of the $f_s$ fluctuations  shown by the KC model is relatively constant (see panel e), regardless of the value of the control parameter $\sigma_p$. These observations point to important dynamical differences between the two models, as will be expanded in the next sections.

 \begin{figure}[h]
 \centering
\includegraphics[width=.7\linewidth]{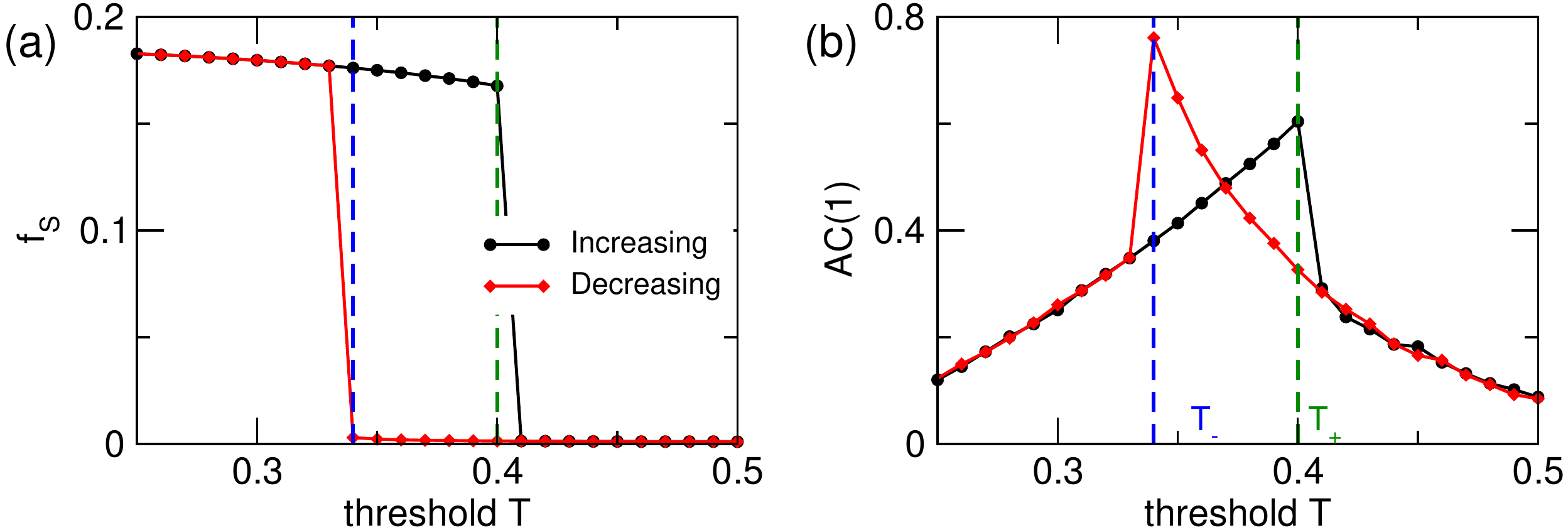}
\caption{{Evolution of the order parameter $f_s$ and its autocorrelation $AC(1)$ as a function of the control parameter $T$ in the GH model}. Panel (a): fraction of active neurons vs.\ threshold $T$. Panel (b): First autocorrelation coefficient $AC(1)$ vs.\ threshold $T$.  In both panels the transition points, $T_-$ and $T_+$, are marked with vertical dashed lines.  Simulations were performed in a network with $\langle k \rangle = 40$ and $\pi=0.6$.  Simulations started at $T_0=0$ and neurons in a random state. $T$ was slowly increased by $\Delta T=0.005$ every $10^5$ time steps up to $T_F=0.9$, then it was decreased back to $T_0$ in the same way. } \label{Fig3}
\end{figure}

\begin{figure}[hb]
\centering
\includegraphics[width=.7\linewidth]{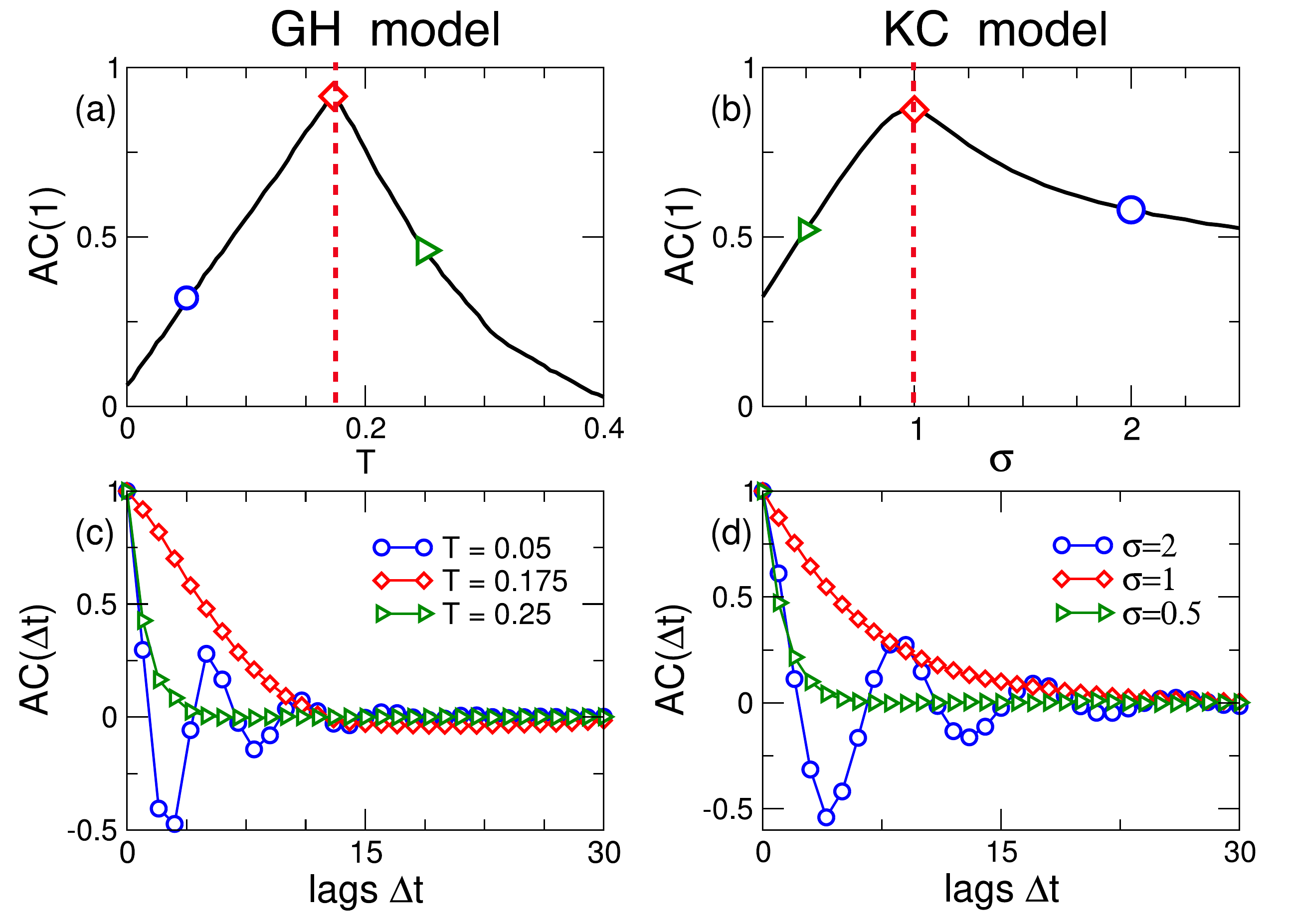}
\caption{Behavior of the first autocorrelation coefficient and the autocorrelation function for network parameters resulting in a continuous phase transition. Top panels: first autocorrelation  coefficient  AC(1) vs.\ the control parameter, $T$ for the GH model (panel a) and $\sigma_p$ for the KC model (panel b).  Bottom panels: Autocorrelation function $AC(\Delta t)$ vs.\ time lag $\Delta t$, for three values of the control parameter, in the supercritical, critical, and subcritical phases, for the GH model (panel c) and the KC model (panel d). Dotted lines in panels (a) and (b) denote the critical point and symbols indicate the values of the control parameters used to compute the data in panels (c) and (d). Network parameters:  $\langle k \rangle  = 10$, $\pi=0.6$. }
\label{Fig4}
\end{figure}
The behavior of the autocorrelation function of the order parameter helps to identify the type of phase transition because it is known to peak near a transition.  We  compute $AC(1)$ for each value of the control parameter, and define $T_+$ as the value that maximizes $AC(1)$ in a run when $T$ is being increased, and $T_-$ as the value that maximizes $AC(1)$ when decreasing $T$. An example for the GH model is presented in Fig. \ref{Fig3}.  For the KC model we defined in the same way $\sigma_p+$ and $\sigma_p-$, although we never observed dis-continuous transitions in that model.

Thus, according to the shape of the curves of $AC(1)$ vs.\ control parameters, we can classify the dynamical behavior: if $AC(1)$ is monotonic, then there is no phase transition, corresponding to the cases in which the network, after a transient goes quiescent. For network topologies in which  $\lvert T_+ -T_-\rvert \ge 2 \Delta T$ (or $\lvert \sigma_+ - \sigma_-\rvert \ge \Delta \sigma_p$) the transition is considered discontinuous, and continuous otherwise.  In other words, after exploring a reasonable range of values of the control parameter, the existence of a maximum in the $AC(1)$ curve indicates (under the present context) a phase transition, which is considered continuous if there is no noticeable hysteresis or discontinuous otherwise.

An example of the behavior of $AC(1)$ in the case of a continuous transition is shown in Fig. \ref{Fig4}. This type of transition is observed in both models for  a wide range of  $\langle k \rangle$ and  $\pi$ values, as will be described in the next section. It can be seen that a change of the control parameter  on a range of values near the critical point is reflected on a non-monotonic change of the $AC(1)$. The plots in the bottom panels illustrate the typical autocorrelation function of the order parameter $f_s$. For control parameter values larger than $T_c$ (or smaller than $\sigma_c$) the activity correlation vanishes quickly as indicated by the green triangle data points.  In the other extreme, for control parameter values smaller than $T_c$ (or larger than $\sigma_c$, i.e.,  data points plotted as blue circles) the function shows an oscillatory pattern. The first zero crossing of the function is dictated by the duration of the refractory period of the neuronal models which is one of the determinants of the collective oscillation frequency. Finally, for values sufficiently close to $T_c$ (or $\sigma_c$) the function $AC(\Delta t)$ decays very slowly (as a power law) as shown in the figure by the data points plotted with red squares.

\subsection{Models's dynamical repertoire on parameter space }

Here we describe the results of a  systematic exploration  of the collective dynamic as a function of network topology in each model.  For each value of  $\langle k \rangle$  and $\pi$, we computed 5 realizations of Watts-Strogatz graphs. In each case we classified the regimes as a function of the control parameter, according to the behavior of $AC(1)$ as explained above.  The dynamical regimes found include \emph{transients to no-activity}, \emph{continuous phase transition} or \emph{discontinuous phase transition (from no-activity to collective oscillations)}.

\begin{figure}[h]
\centering
\includegraphics[width=0.7\linewidth]{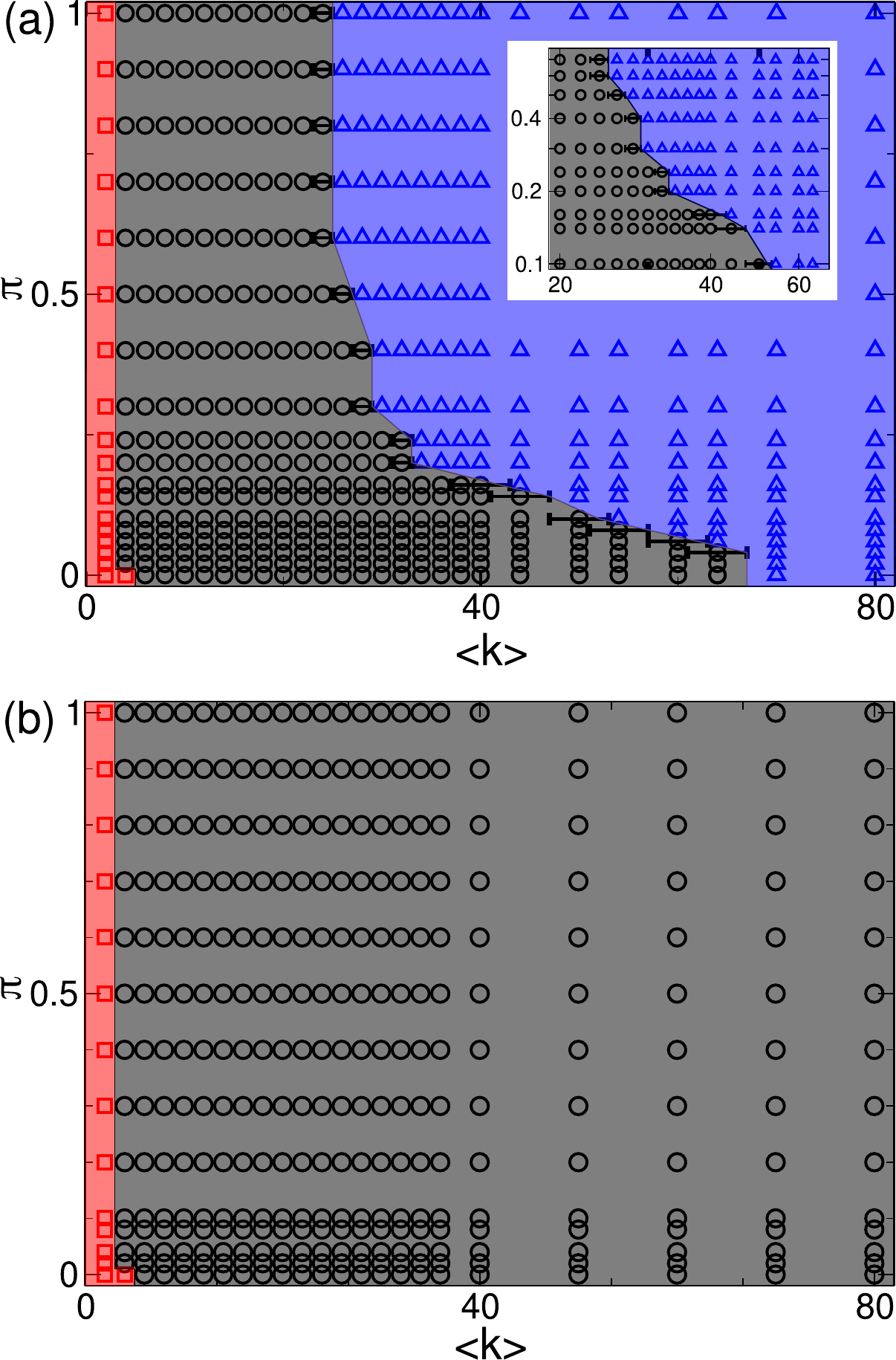}
\caption{Classes of collective dynamics emerging at different network topologies for both models (GH model in panel a, KC model in b). The graphs summarize the dynamical regimes observed for a wide range of network parameters (average connectivity $\langle k\rangle$ and degree of disorder $\pi$).
Blue regions (triangles denote the values tested) indicate those values for which the networks exhibited discontinuous phase transitions to collective oscillations. Black regions (with  circles) indicate continuous phase transitions, and red regions (with squares) transient dynamics to inactivity but no phase transition. The inset in (a) depicts a portion of the same data  plot in the main panel in log-log scale. }\label{FigMap1}
\end{figure}

The results in Fig.~\ref{FigMap1} show the regions of parameters at which each regime was observed. In brief, both models exhibit no-transition for network topologies with $\langle k \rangle  = 2$   and connectivity disorder  $\pi > 0$ (red zone with squares in Fig.~\ref{FigMap1}). For $\pi=0$ the same regime extends to $\langle k \rangle < 6 $ in both models.

For networks with relatively  large values of $\langle k \rangle$ both models exhibit a \emph{continuous phase transition}  as in the example featured already in Fig.~\ref{Fig4}. (black zone with circles in Fig.~\ref{FigMap1}). The main difference between the models is found for relatively high values of degree and disorder. At this region of parameters the GH model shows a  discontinuous phase transition (blue zone  with triangles), while the KC model a continuous one.

\begin{figure}[h]
\centering
\includegraphics[width=0.7\linewidth]{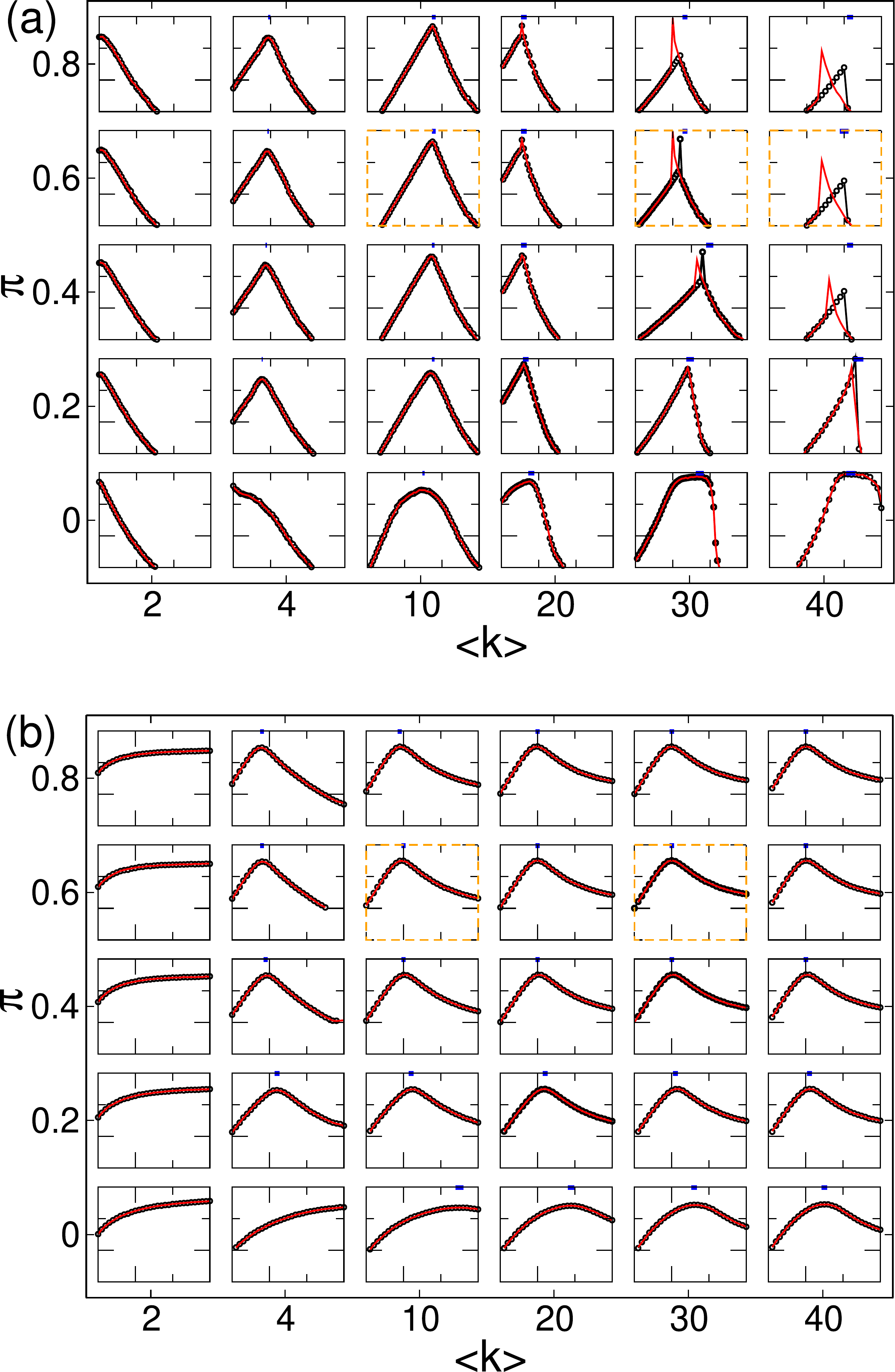}
\caption{Representative examples of the typical behavior of AC(1) as a function of the control parameter for diverse topologies in both models  (GH model in panel a, KC model in b).  Simulations where run for $5\cdot 10^4$ time steps for each value of $T$ or $\sigma_p$, increasing (black circles) or decreasing (red lines) by $\Delta T= 0.005$ or $\Delta\sigma=0.05$. For the GH model, $x$-axis range is  $[0:0.3]$, for $\langle k \rangle=2$, 4 and 10,  and  $[0.2:0.5]$ for other $\langle k \rangle$ values.  For the KC model,  $x$-axis range is $\sigma_p=0.5$ to $\sigma_p=2$.  $y$-axis range is $AC(1)=0.25$ to $AC(1)=1$ for both models. A thick blue dash over the upper x-axis marks a $\pm 3 \%$ range of the control parameter, about it's critical value (or   $T_+$ for discontinuous transitions). The boxes remarked with dashed lines correspond to the parameters used in Figs.~\ref{Fig2}-\ref{Fig3}-\ref{Fig4}. }\label{FigMap2}
\end{figure}

The results in Fig.~\ref{FigMap2} are representative examples of the behavior of $AC(1)$ as a function of the control parameter for selected values of $\langle k \rangle$ and $\pi$.  For the GH model, the largest values of $\langle k \rangle$ and $\pi$ show clear hysteresis, with the peaks for the case of increasing $T$ at a higher value than the peak found when is decreased.  In most cases, the increasing and decreasing sweeps of control parameter yield the same curve, with a maximum value of $AC(1)$ close to $1$.  For $\pi=0$,  the (single) peak tends to be rather broad. Finally, for $\langle k \rangle=2$ and any value of $\pi$,  and for $k=4$, $\pi=0$, $AC(1)$  behaves monotonically, which is indicative of no phase transition.  The KC model shows less variation among the curves, with only a narrow range of monotonous curves, and most of the $\langle k\rangle$, $\pi$ plane yielding continuous transitions.

\begin{figure}[h]
\centering
\includegraphics[width=0.7\linewidth]{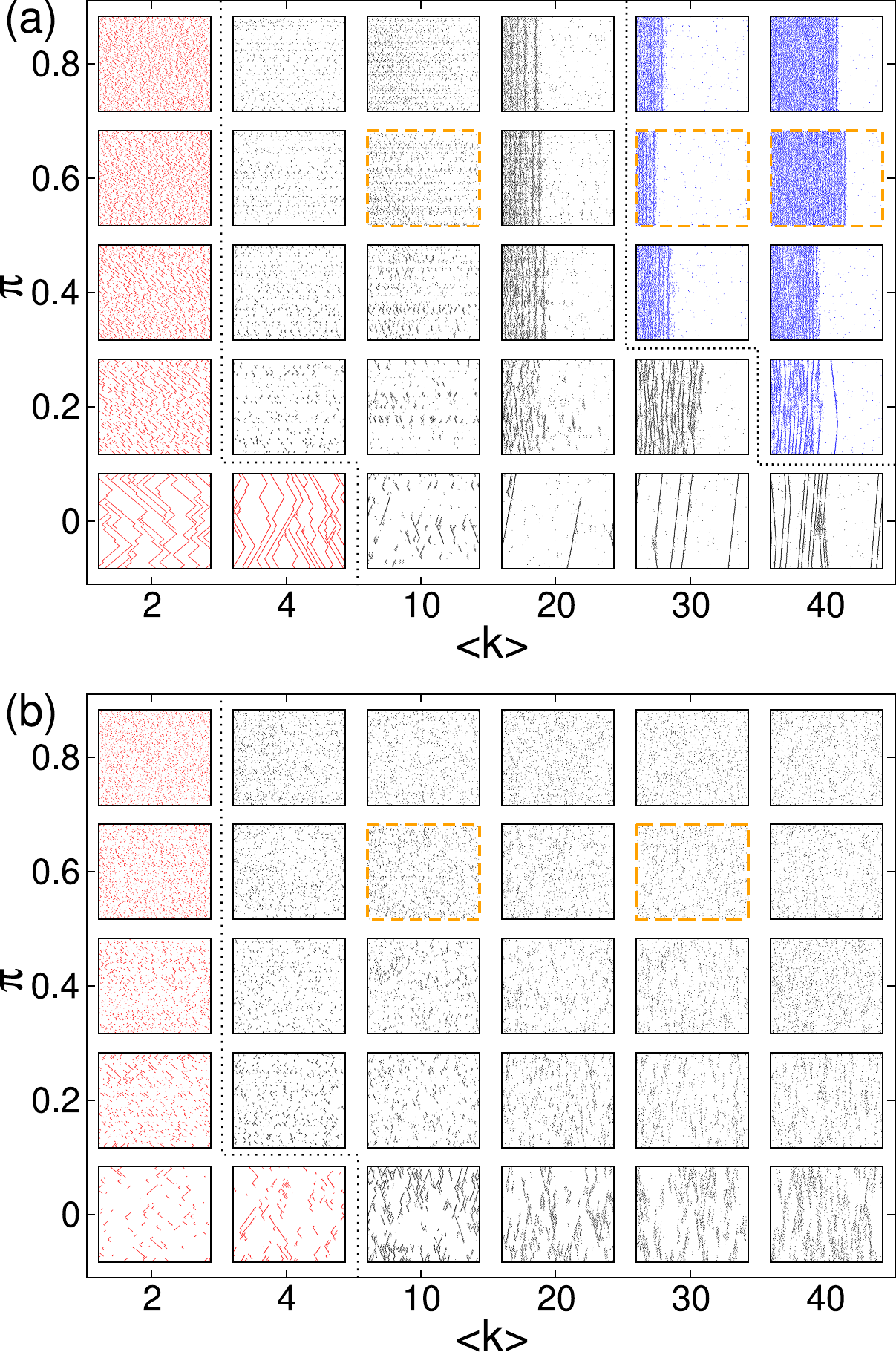}
\caption{Typical rasters of activity for networks with diverse topologies, while varying the control parameter around the critical value, for both models (GH model in panel a, KC model in b). The dots in each box denote activity of a subset of 300 neurons (ordered in the $y$ axis), for 300 time steps ($x$ axis), as the control parameter is changed continuously about the critical point. Other network parameters as in Fig. \ref{FigMap2}. Dot color matches the regions of Fig.~\ref{FigMap1}. The dotted black lines indicate the limits of the  parameters that exhibit   phase transitions.  The boxes remarked with dashed lines correspond to the parameters used in Figs.~\ref{Fig2}-\ref{Fig3}-\ref{Fig4}.}\label{FigMap3}
\end{figure}

Finally, the results on Fig.~\ref{FigMap3} show examples of the spiking patterns observed as the control parameter is increased, for the same selected  values of $\langle k \rangle$ and $\pi$ illustrated in Fig.~\ref{FigMap2}. The patterns were obtained by varying the control parameters from $3\%$ below $T_c$ (or $T_+$ in the discontinuous case) to $3\%$ above (or from $3\%$ above to $3\%$  below the critical value of $\sigma_p$ for KC model),  and recording the spikes of 300 neurons along 300 time steps. We have used $T=0$ or $\sigma=2$ for networks showing no phase transition. For the GH model there are several cases (blue raster plots) of discontinuous transitions where there is a sharp decrease of activity after crossing $T_+$.  Continuous transitions with large $\langle k \rangle$ (such as $\langle k \rangle=20$,  $\pi\geq 0.4$), show bursts of synchronized activity that disappear for $T$  slightly above $T_c$ (black raster plots).  For $\pi=0$ the network topology corresponds to a circle (or to a torus for larger $\langle k \rangle$ values), so that neurons spiking at time $t+1$ are close neighbors of those spiking at time $t$, leading to linear wave-like propagation.

\section{Discussion}
\label{Discussion}

Summarizing, we have revisited two simplified models of neuronal activation to show that subtle differences may result in very different collective dynamics  when embedded on networks. We found that the KC model dynamical repertoire includes, as a function of its control parameter, only a continuous phase transitions being, by design, insensitive to the network topology. This is at odds with the GH model, in which each neuron outcome is influenced by its connectivity degree and therefore by the overall network topology.

We have used the first autocorrelation coefficient $AC(1)$ of the order parameter fluctuations and the presence or absence of hysteresis to identify whether a dynamic transition is present, and to distinguish continuous from discontinuous transitions. This observable is sensitive enough to even being able to tune a system towards criticality \cite{Chialvo2020Control}. None of the present results depend on the use of the autocorrelation to track the dynamics. The presence of  phase transitions and hysteresis in these  models has been studied with other observables, such as the fraction of active sites $f_S$, the variance of activity fluctuations $\sigma_{f_S}^2$, or cluster quantities such as  the size of the largest or the second largest cluster ($S_1$, or $S_2$) as in \cite{Zarepour,Haimovici2013}, yielding similar results.  We used $AC(1)$ because its computation is straightforward and easy to replicate, it is almost parameter free, therefore very convenient for comparing the two models.

The key difference between the two models is in the rule that determines how the activity propagates from a given neuron to its connected neighbors.  The GH model mimics a discrete integrate-and-fire process taking place in real neurons. There,  the ``decision'' to fire is post-synaptic, based on the amount of total depolarization, on a small patch of membrane, produced by  the contribution of hundreds to thousands of impinging neurons (notice the sum in Eq.~\eqref{eq:act-modA}). Disregarding the spontaneous activation term, the rule in the GH model is completely deterministic.  In fact it is equivalent to a discretized partial differential equation, where the state $S_{I}(t+1)$ of a post-synaptic neuron $I$ at time $t+1$ depends only on local quantities: it is determined by the previous state of that neuron, $S_I(t)$, the contribution of all other presynaptic neurons  $J_1$ to $J_k$ at  time $t$, the weights $W_{ij}$ of the connections, and the excitation threshold, i.e., $S_{I}(t+1)= f\left[ S_{I}(t), S_{J_1}(t),\ldots,S_{J_k}(t), W_{J_{i=1..k},J},T\right]$.

In contrast, in the KC model the propagation rule is a probabilistic (notice the product Eq. \eqref{eq:act-modB})  contagion-like process \cite{Branching}, where a single excited neuron determines, according to a prescribed value of $\sigma_p$ how many of all of the neurons that connects to will fire next time. Thus, here the decision of how many neurons will be activated is presynaptic:  a spiking neuron $J$ excites on average $\sigma_p$ post-synaptic $I$ neurons, independently of the state of the other neurons connected to the same $I$ neuron.  Since $\sigma_p$ accounts already for the average out degree of the network, the rule in Eq. \eqref{eq:act-modB} determines, \emph{for each $J$ neuron independently} the probability that such active neuron will have no-descendants, one or more than one descendants.  It is then rather unsurprising that the KC model is insensitive to the network topology. 

Although this is not the focus of the present report, it is worth to mention that the KC model rule is biologically implausible.  As explained above, the actual activation mechanism of a given neuron involves a myriad of  influences (thousands in mammalian brains) over a very small area. In addition, one has to consider that the output of an active neuron, after propagating trough its axon from hundreds  of microns to millimeters,  will stimulate all its contacts roughly \emph{equally}.  Thus, a neuron in order to excite a given number of neighbors would actually have to have information on the number and state (active, refractory or silent) of the post-synaptic neurons and also for other presynaptic neurons attempting to excite the same neuron. This is biologically highly unrealistic, because the neurons involved may be centimeters away from each other, without direct connection among them. Moreover, such hypothetically very well informed $J$ neuron will have to selectively cancel its stimulation strength with certain $I$ neurons, as dictated by the value of $\sigma_p$, a requirement completely impossible for biological neurons.

Of course, the discussion above is not affecting the valuable points made in ref.~\cite{Kinouchi2006} where the KC model was introduced  to show that a network of identical excitable elements achieves maximum dynamic range at criticality. Since the simulations in Ref.~\cite{Kinouchi2006}  where made on fixed topology networks (Erd\"os-Renyi) the present results are not affecting any of its conclusions which, is worth noticing, were replicated in many other models as well as in experiments.
Less clear are the results in other cases, when the KC model was used to study the dynamics of non random topologies. These include the deviations from mean field behavior found in scale free networks, as reported by Copelli \& Campos \cite{CopelliCampos2007} or  Mosqueiro \& Maia \cite{Mosqueiro}. We note in passing that there are multiple instances in which  the KC model was (mis)named as ``Greenberg \& Hastings stochastic model'', somewhat confusing according to  the present results, as in the reports by Copelli \& Campos \cite{CopelliCampos2007}, Wu \emph{et al.}, \cite{wu2007}, Asis \& Copelli \cite{AssisCopelli2008}, Mosqueiro \& Maia \cite{Mosqueiro}, to name only a few. 

It is worth to note that the crucial influence of topology on the type of dynamics exhibited by excitable models has been discussed earlier by Kuperman \& Abramson \cite{Abrahamson2001Epidemiological} in the context of epidemics.  They found that the network degree and disorder determines the conditions at which endemic or epidemic situations occur.

To conclude, the point here is not that one needs super-realistic microscopic rules to build a valid model, but that microscopic rules do matter, and may lead to different collective behavior.  In particular, we have shown that the KC model exhibits only one type of transition, independent of a large variation in the network topology.  The observation that topology has influence on the dynamics of certain models is fundamental at the present time, where several large scale  international scientific collaborations are devoted to map and study the consequences of features of the human brain connectome \cite{markram2006blue,alivisatos2012brain,sunkin2012allen}. 
Examples include the numerical simulations using simplified models  over derived connectomes, in order to understand brain functioning \cite{odor2016critical}. However, \emph{not all neural models are the same}, and special care should be taken on the biases and limitations introduced by  the applied models, before drawing conclusions on real brains.

\section{Acknowledgements}

Work supported by the BRAIN initiative Grant U19 NS107464-01. DAM acknowledges financial support from ANPCyT Grant PICT-2016-3874 (AR) and the use of computational resources of the IFIMAR (UNMdP-CONICET) cluster.


\title{Supplemental material for: ``Apparently similar neuronal dynamics may lead to different collective repertoire'' by Sánchez Díaz \emph{et al.}}

\maketitle

Here we comment on the implementation of the models used to generate the results in the main text. 
The respective codes in Fortran 90 and Python 3 can be found on  GitHub\cite{codes}.

\section{ Implementation of the Watts-Strogatz Network}

In Fortran 90, the code \verb|Matrix.f90| is used to generate  an undirected Watts-Strogatz network with parameters $N$, $\pi$, and $\langle k\rangle$ which is saved as a two columns  ascii file \verb|MyMatrix.txt| of length $N\cdot \langle k\rangle/2$. Each row has two integer numbers, $i$ and $j$, which denote a symmetric edge between nodes  $i$ and $j$.

For Python, the Watts-Strogatz network is generated  using the available code from the \verb|networkx| package.

\section{Implementation of the Greenberg \& Hastings (GH) Model}
\subsection{Fortran 90 code}
The dynamics of the GH model,  given by Eqs. 1 (a)-(c) of the main text, is implemented by the program \verb|GH_model.f90|.

The main code reads the file \verb|MyMatrix.txt| (generated as explained in the previous section), and assigns weights to each connection  according to an exponential distribution as explained in the main text. For each value of the control parameter $T$, it runs the \verb|ConstantStep| subroutine for a given number of steps (typically $50.000$), updating synchronously the network state using the \verb|Step| subroutine. At the end of the \verb|ConstantStep| subroutine,  it computes the order parameter $f_S$ and  the autocorrelation coefficient  $AC(1)$.
 At each time step $t$ the \verb|Step| subroutine computes the state of all neurons at time $t+1$, $S(t+1)$, from $S(t)$. For each neuron $i$:

\begin{enumerate}
 \item If the neuron $i$ was \emph{quiescent} (i.e., $S_i(t)=0$),  it will become \emph{active} $S_i(t+1)=1$ with probability $r_1$. If that does not happen, a loop over all $j$ connected to $i$ is made. For each active $j$  (i.e., $S_j(t)=1$), the weight $W_{i,j}$ is summed up. If the sum is larger than $T$, then $S_i(t+1)=1$.
 
 \item If the neuron was \emph{active}, (i.e., $S_i(t)=1$), it will become \emph{refractory} (i.e., $S_i(t+1)=2$) always.
 
 \item If the neuron was \emph{refractory}, (i.e., $S_i(t)=2$), it will become \emph{quiescent} (i.e., $S_i(t+1)=0$) with probability $r_2$.
\end{enumerate}

 All neurons are updated synchronously.

\subsection{Python 3 code}

The code \verb|GH_model.py| in Python 3  follows the same steps and uses the same names for variables as in the  Fortran 90 code. The code requires the following packages: \verb|numpy|, \verb|random| and \verb|networkx|.

The neural activity as a function of time is saved on files named \verb|actN(N)K(K)PI(100*pi)T(1000*T).txt|,  
 where the value of $\pi$ is multiplied by 100 and the value of $T$ is multiplied by 1000, so that they are represented by integer numbers. For instance, for parameters $N=5000$, $\langle k\rangle=10$, $\pi=0.6$ and $T=0.2$ the output file will be named \verb|actN5000K10PI60T200.txt|

\section{Implementation of the Kinouchi \& Copelli (KC) Model}

In the following we describe the implementation of the code for the KC model used in the main  manuscript. Two possible variations of the model are described later in section \ref{alternative}. The KC code follows the same structure as the GH counterpart, starting by reading the file \verb|MyMatrix.txt| but changing a few relevant parts: the \verb|Step| subroutine is  obviously different and the program cycles trough values of $\sigma_p$ instead of $T$.

Here we call $KC_m$ to the code implementation used for the results described in the main text,  which follows Eqs. 2 (a)-(d). The codes are \verb|KC_m.f90| and  \verb|KC_m.py| for Fortran 90 and Python 3 respectively. 

 At each time step $t$ the \verb|Step| subroutine computes the state of all neurons at time $t+1$, $S(t+1)$, from $S(t)$. For each neuron the \verb|Step| subroutine updates the states as follows:

\begin{enumerate}

 \item For each active neuron $j$ (i.e., $S_j(t)=1$) a loop over all output connections is made.  Each \emph{quiescent} neuron $i$  (i.e., $S_i(t)=0$), will become \emph{active} with probability  $\sigma_p W_{ji}$,

 \item If the neuron $j$  was \emph{quiescent} (i.e., $S_j(t)=0$),  it will become \emph{active} ($S_j(t+1)=1$) with probability $r_1$, 
 
 \item If the neuron $j$ was \emph{active}, (i.e., $S_j(t)=1$), it will become \emph{refractory} (i.e., $S_j(t+1)=2$) always.

 \item If the neuron $j$ was \emph{refractory}, in state $m$, (i.e., $S_j(t)=m>1$), it will become \emph{refractory} with state $m+1$ (i.e., $S_j(t+1)=m+1$) unless $m=n$. In that case, if becomes   \emph{quiescent} (i.e., $S_j(t+1)=0$).
\end{enumerate}

All neurons are updated simultaneously. Within this implementation, it may happen that, in the same time step more than one active neuron activates the same quiescent neuron.  

\section{Two alternative implementations of the KC model.}
\label{alternative}

As mentioned in the previous section, we explored some algorithmic  alternatives for the implementation of the KC model. We will describe two of them here.
\subsection{Backward update implementation}

What we call the \emph{backward} implementation (see codes \verb|KC_b.f90| and  \verb|KC_b.py|) is described also by Eqs. 2 (a)-(d) of the main text. The only  subtle difference with $KC_m$ is in the activation loop within the \verb|Step| subroutine: now it is performed not over the \emph{active} neurons but over the \emph{quiescent} neurons. Therefore in this implementation,  the \verb|Step| subroutine is:  

\begin{enumerate}

\item If  neuron $i$  was \emph{quiescent}  (i.e., $S_i(t)=0$), a loop  over all input
 connections $j$  is made. For each \emph{active} neuron $j$  (i.e., $S_j(t)=1$), an  attempt to \emph{activate} $i$ with probability   $\sigma_p W_{ji}$ is made.

\item If the neuron $j$  was \emph{quiescent} (i.e., $S_j(t)=0$),  it will become \emph{active} ($S_j(t+1)=1$) with probability $r_1$, 
 
 \item If the neuron $j$ was \emph{active}, (i.e., $S_j(t)=1$), it will become \emph{refractory} (i.e., $S_j(t+1)=2$) always.

 \item If the neuron $j$ was \emph{refractory}, in state $m$, (i.e., $S_j(t)=m>1$), it will become \emph{refractory} with state $m+1$ (i.e., $S_j(t+1)=m+1$) unless $m=n$. In that case, if becomes   \emph{quiescent} (i.e., $S_j(t+1)=0$).
\end{enumerate}

Similar to the $KC_m$ implementation, it may happen that neuron $i$ is simultaneously activated by several $j$'s.

\begin{figure}[hb]
\centering
\includegraphics[width=0.65\linewidth]{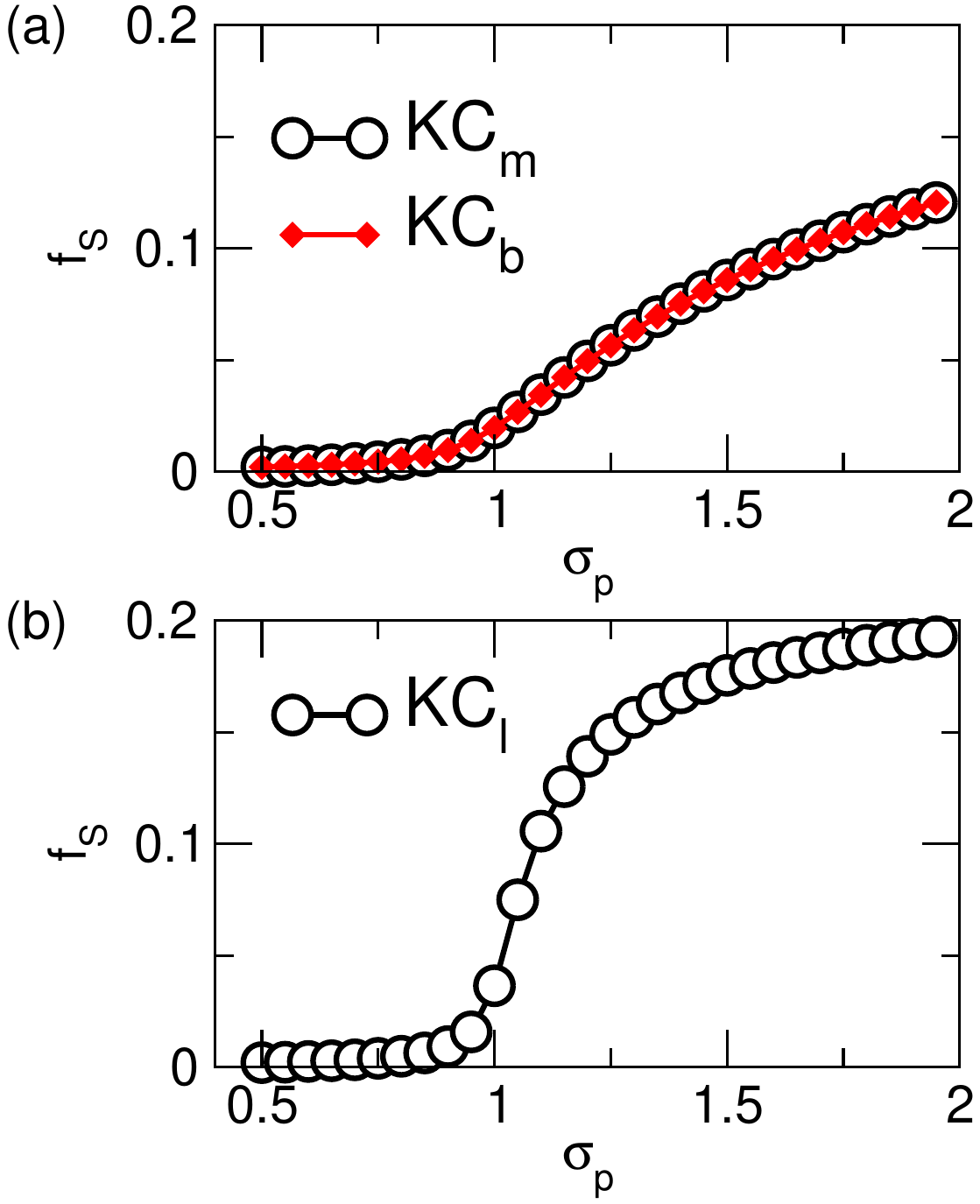}
\caption{Comparison of alternative implementations of the KC model. The fraction of active neurons, $f_s$ as a function of the order parameter $\sigma_p$, using the  $KC_m$ (black circles), and the  $KC_b$ (red diamonds), models are shown in (a). $KC_l$  results are shown in (b). In all the cases, the Fortran 90 codes, with  $5\cdot 10^4$ time steps and $\Delta \sigma_p=0.05$ were used. A Watts-Strogatz network with parameters: $N=20000$, $\langle k \rangle=10$, $\pi=0.6$ was used.  }\label{FigSMCode}
\end{figure}

 An example of $f_S$  as a function of $\sigma_p$ computed with both  $KC_b$ and $KC_m$  implementations in  Fortran 90, is shown in Fig. \ref{FigSMCode} (a). The results are identical, nevertheless, one implementation may be considerably faster, due to the fact that, depending on the case, all or only the active neurons need to be updated. For instance, using \verb|gfortran|, compiled with \verb|-O3| flag, the $KC_m$ implementation takes about $2/3$ of the time required by the $KC_b$ implementation to  reproduce Fig. \ref{FigSMCode} (a).

 \subsection{Local out-degree implementation}

 We have also explored another implementation of the KC model, we called $KC_l$, by slightly modifying its activation rule. In this case, in each update the activation probabilities of each neuron depends on its out-degree (and the state of its target neurons). Specifically, a given active neuron $j$ counts how many connected quiescent neurons has at time $t$ (i.e., $\sum_{m=1}^{k_{out,j}}  \delta_{S_{m},0}$), and re-normalizes its contribution among them.

 All the other transition rules and parameters are the same as in the $KC_m$ model. 
 In this case the \verb|Step| routine reads:

 \begin{enumerate}

  \item For each active neuron $j$ (i.e., $S_j(t)=1$) a loop over all output connections is made.  
The number of $i$ neurons connected with $j$ that are \emph{quiescent} is counted and defined as $NC$.  Each \emph{quiescent} neuron $i$, will become \emph{active} with probability   $2\sigma_p W_{i,j}/NC$.

 \item If the neuron $j$  was \emph{quiescent} (i.e., $S_j(t)=0$),  it will become \emph{active} ($S_j(t+1)=1$) with probability $r_1$, 
 
 \item If the neuron $j$ was \emph{active}, (i.e., $S_j(t)=1$), it will become \emph{refractory} (i.e., $S_j(t+1)=2$) always.

 \item If the neuron $j$ was \emph{refractory}, in state $m$, (i.e., $S_j(t)=m>1$), it will become \emph{refractory} with state $m+1$ (i.e., $S_j(t+1)=m+1$) unless $m=n$. In that case, if becomes   \emph{quiescent} (i.e., $S_j(t+1)=0$).

\end{enumerate}

\begin{figure}[h]
\centering
\includegraphics[width=0.95\linewidth]{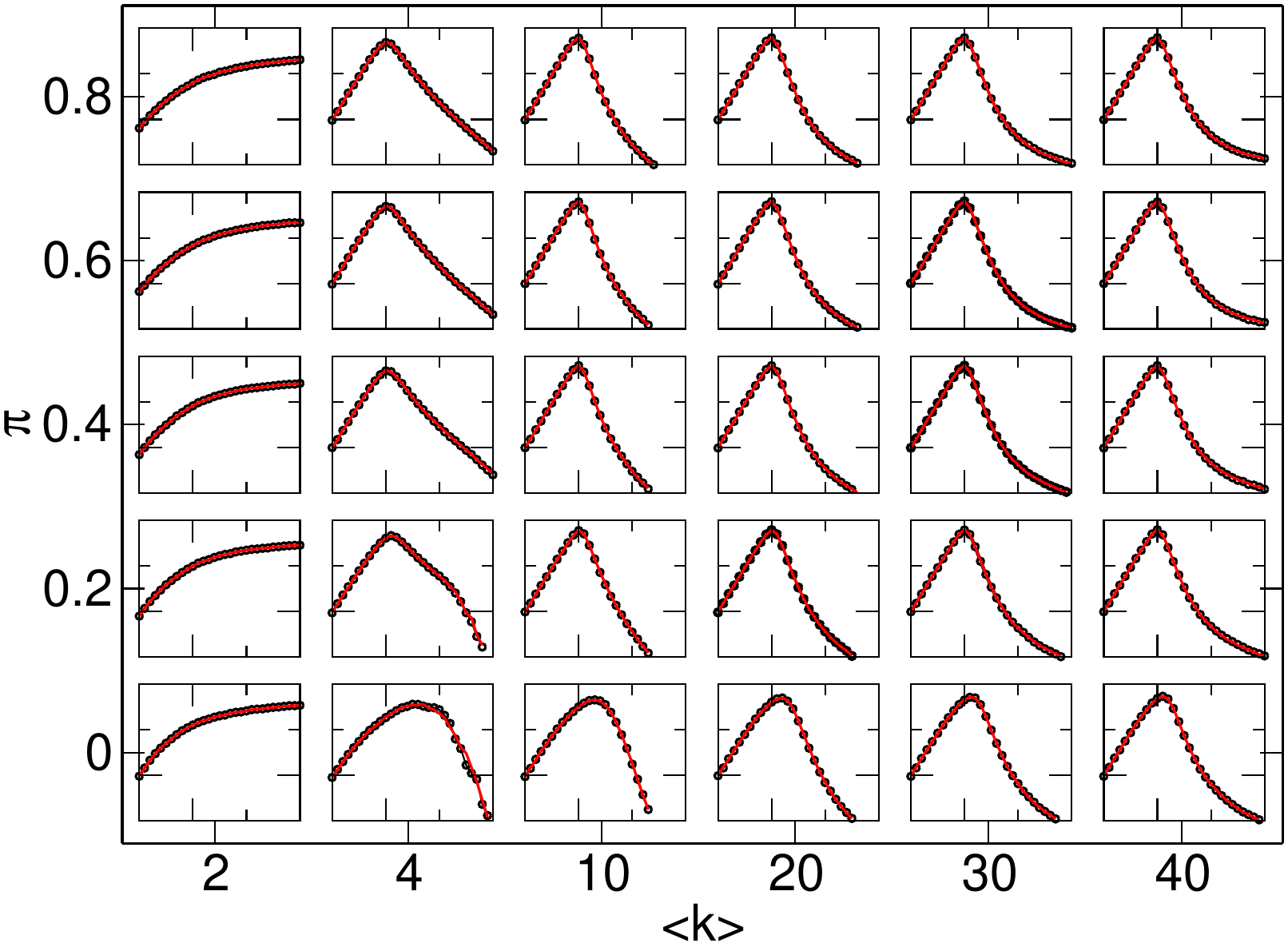}
\caption{Results obtained with the alternative $KC_l$ described in section IVb. The data shows representative examples of the typical behavior of $AC(1)$ as a function of $\sigma_p$ for diverse topologies.  Simulations were run  increasing (black circles) or decreasing (red lines)  $\sigma_p$ values by $\Delta\sigma=0.05$ every $5\cdot 10^4$ time steps. In each box, the  $x$-axis range is $\sigma_p=0.5$ to $\sigma_p=2$ and the $y$-axis range is $AC(1)=0.25$ to $AC(1)=1$.  }\label{FigMapSM}
\end{figure}

 $KC_l$  differs from $KC_{m/b}$ in that, in  $KC_l$, the probability that a given neuron $j$ activates neuron $i$  depends of its degree (the number of connections of $j$), and the state of all other output connections. This introduces differences that deserve to be further explored.  An example of  $f_S$ as a function $\sigma_p$, computed with Fortran 90 codes is shown in Fig. \ref{FigSMCode} (b). Further results for $AC(1)$ as a function of $\sigma_p$ for $KC_l$ (similar to Fig. 6-(b) on main text), are shown in Fig.~\ref{FigMapSM}. Similar to  $KC_m$, $KC_l$ model does not present discontinuous transitions: for $\langle k \rangle =2$, there are \emph{transients to no-activity}, while for larger values of  $\langle k \rangle$, there are  \emph{continuous phase transitions}. Notice that  for this model the $AC(1)$ decays relatively faster for $\sigma_p$ values away from the critical value $\sigma_p=1$ (compare with Fig. 6 on main text).

\end{document}